\documentclass[12pt,twoside,a4paper]{article}
\usepackage{epsfig}
\usepackage{setspace}

\def\beq{\begin{equation}}
\def\eeq#1{\label{#1}\end{equation}}
\def\eeqn{\end{equation}}

\def\beqa{\begin{eqnarray}}
\def\eeqa#1{\label{#1}\end{eqnarray}}
\def\eeqan{\end{eqnarray}}

\let\bar=\overbar

\newcommand{\captionfonts}{\footnotesize}

\makeatletter  
\long\def\@makecaption#1#2{%
  \vskip\abovecaptionskip
  \sbox\@tempboxa{{\captionfonts #1: #2}}%
  \ifdim \wd\@tempboxa >\hsize
    {\captionfonts #1: #2\par}
  \else
    \hbox to\hsize{\hfil\box\@tempboxa\hfil}%
  \fi
  \vskip\belowcaptionskip}
\makeatother   

\def\Title#1{

\setbox0\vbox{\vskip-70pt
{\scriptsize{
{\em
\begin{raggedright}
Primer Encuentro de la Radioastronom{\'{\i}}a Espa{\~n}ola\\
~~~~~~~~~~~~~~~``Memorial Lucas Lara''\\ 
J.C.~Guirado, I.~Mart{\'{\i}}-Vidal, and J.M.~Marcaide (eds.)\\ 
May 9th-11th 2006, Valencia, Spain\\
\end{raggedright}
}}}
}

\wd0=0pt
\ht0=0pt
\box0

\begin{doublespace}
\vspace*{-25pt}
\begin{center} {\Large {\bf #1} } \end{center}
\end{doublespace}

\medskip

}

\newcommand{\runninghead}[2]
{\markboth{\small{\it {#1}}}{\small{\it{#2}}}}

\def\Dslash{\not{\hbox{\kern-4pt $D$}}}
\def\dslash{\not{\hbox{\kern-2pt $\del$}}}

\def\msb{{\bar{\ssstyle M \kern -1pt S}}}

\def\Dm2{\Delta m^{2}}
\def\eV2{\mbox{eV}^{2}}
\def\B8{^{8}\mbox{B}}
\def\Be7{^{7}\mbox{Be}}

\begin{document}


\Title{VLBI DIAGNOSTICS OF JET INSTABILITIES IN 0836+710}

\begin{raggedright}

M.~Perucho, A.~Lobanov\\

{\small{\it Max-Planck-Institut f\"ur Radioastronomie, Auf dem
H\"ugel 69, 53121, Bonn, GERMANY}} \\

\runninghead {M.~Perucho \& A.~Lobanov} {VLBI diagnostics of jet
instabilities in 0836+710}

\thispagestyle{empty}
\medskip\smallskip


{\small {\bf Abstract:} In this paper, we present new
VLBA\footnote{Very Long Baseline Array, operated by the National
Radio Astronomy Observatory, USA.} observations of the radio jet
in the quasar S5\,0836+710 at 8 and 22 GHz. The identification of
the ridge lines allow us to interpret the jet structure in terms
of Kelvin-Helmholtz instability. Combined with previous epochs of
VLBA and VSOP\footnote{VLBI Space observatory Program led by the
Japanese Institute at Space and Astronomical Science.} data at 1.6
and 5 GHz, these new observations will allow us to study the
evolution of the instabilities in the jet. We have detected
signatures of possible jet disruption in the jet due to the growth
of instabilities, which points towards a possible morphological
classification of the source as an FRI, contrary to what
previously thought in terms of luminosity criteria.}

\end{raggedright}
\begin{singlespace}


\section{Introduction}

The quasar S5\,0836+710 is at a redshift of 2.16. At kiloparsec
scales it was classified as a FRII source following luminosity
criteria for a secondary component, located at 2$^{\prime\prime}$
from the core. Its luminosity ($P_{21cm}=10^{27}$ W/Hz, O'Dea et
al. \cite{odea}) is clearly over the threshold
($P_{21cm}=10^{24.5}$ W/Hz) separating FRI and FRII sources.
However, Hummel et al. \cite{hum} observed a significant loss of
collimation in the extended jet, contrary to the morphology
expected for an FRII source. At parsec scales, a link was found
between the ejection of a superluminal component (component B3)
and a gamma-ray-optical flare (Otterbein et al. \cite{otte}). The
spectral evolution of the component B3 is consistent with the
synchrotron self-absorption and adiabatic expansion in a
synchrotron emitting plasma condensation. A Lorentz factor
$\gamma=12$ and a viewing angle of $\theta=3^\circ$ were deduced
for this component. Otterbein et al. \cite{otte} also found
several kinks in the jet and suggested that an apparent helical
morphology at the jet can be due to growing MHD instabilities in
the relativistic plasma.

Lobanov et al. \cite{lo98}, \cite{lo06} presented VLBA and VSOP
observations of 0836+710 at 1.6 GHz and 5 GHz, and interpreted the
helical structure in the jet as a result of growing
Kelvin-Helmholtz instability. Using the linear stability theory
they derived the Mach number of the jet ($M_j=6$) and the density
ratio between the jet and the external medium
($\rho_j/\rho_a=0.04$).

In this paper, we present new VLBA observations of 0836+710 at 8
and 22 GHz at two different epochs separated by 16 months (July
1998 and November 1999) and combine them with the previous
observations of this object by Lobanov et al. \cite{lo98},
\cite{lo06}. The aim of this work is to obtain information about
the jet morphology on scales of 2-200 mas, and study structural
changes in the radio jet on timescales of $\sim 2$ years.

\section{Observations}
\begin{figure}[!htb]
\begin{center}
\epsfig{file=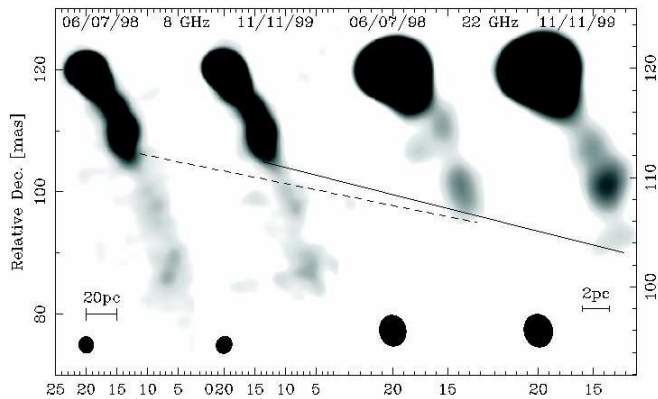,width=0.7\textwidth} \caption{VLBA
images of the jet in 0836+710 at 8 and 22 GHz for two different
epochs. Lines indicate the extent of the 22 GHz jets in the 8 GHz
images. Structures few mas long are observed. Both frequencies are
tapered in order to reveal extended emission.} \label{fig:f1}
\end{center}
\end{figure}

In Fig.~\ref{fig:f1}, we show the maps from the two epochs of VLBA
observations at 8 and 22 GHz. We can see structures in the jet
from a few milliarcseconds to several tenths of milliarcseconds.
Combined with previous observations (Lobanov et al. \cite{lo98},
\cite{lo06}), these data enable to identify structures on scales
from 2 mas (22 GHz) to 200 mas (1.6 GHz).

\begin{figure}[!htb]
\begin{center}
\epsfig{file=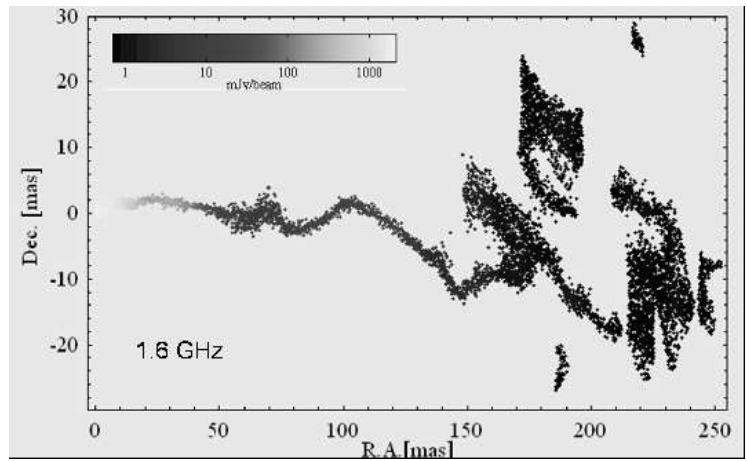,width=0.7\textwidth} \caption{Ridge
line of the jet in 0836+710. The ridge line is measured from the
1.6 GHz VLBA image (Lobanov et al. \cite{lo06}). The grey scale
shows the flux density of the peak. A helical pattern with a
$\sim$100 mas wavelength is observed. After about two wavelengths
the jet is no more visible, either because of dimming, or due to
disruption.} \label{fig:f2}
\end{center}
\end{figure}

In order to measure the wavelengths of the perturbations
developing in the jet, we identify the ridge line of the jet at
different frequencies. The ridge line is traced by maxima of
emission in brightness profiles taken across the jet. In
Fig.~\ref{fig:f2} we plot the locations of these maxima for the
1.6 GHz image from Lobanov et al. \cite{lo06}. The ridge line
shown in Fig.~\ref{fig:f2} reveals a clear helical pattern
extending up to $\sim 150$ mas distance from the jet origin. At
larger distances, this pattern becomes disrupted, possibly
reflecting real disruption of the jet flow. The same kind of plot
for the 8 GHz image from the first of the two epochs presented
here reveals structures on much smaller scales (2-4 mas). A
combination of ridge line measurements made in a range of
frequencies from 1.6 to 22 GHz, will provide reliable verification
of the physical properties of the flow and the ambient medium on
very different scales.

\section{Discussion}
In Fig.~\ref{fig:f2}, we observe a possible signature of
disruption of the flow at distances $> 150$ mas. The fact that the
jet becomes decollimated at kiloparsec scales and shows an
irregular structure in the region of interaction with the external
medium (Hummel et al. \cite{hum}), points also towards this
possibility. In this scenario, the secondary component observed at
kiloparsec scales could be a subrelativistic remnant of the
disrupted flow advancing through the intergalactic medium. Such
features are commonly seen in numerical simulations (e.g., Perucho
et al. \cite{pe05}, \cite{pe06}, Rosen \& Hardee \cite{roha})

Higher dynamic range observations should enable verifying this
scenario. If it is confirmed, this source, first thought to be an
FRII due to luminosity arguments, would have to be considered as
an FRI object, based on its morphology and evolution. This
\emph{dual} nature would turn 0836+710 into a peculiar object
requiring further studies.

Our plans for this project include application of gaussian fits to
the emission profiles in order to identify the ridge lines at all
epochs and frequencies and use observations at four different
epochs to derive the wave speeds of the observed structures.
Further structural analysis will be performed by applying wavelet
analysis and deriving the characteristic wavelengths and physical
parameters of the jet. These measurements will be confronted by
performing multidimensional RHD/RMHD numerical simulations
(Perucho et al. \cite{pe06}). Finally, new high dynamic range
observations with MERLIN\footnote{Multi Element Radio Linked
Interferometer operated by Jodrell Bank Observatory at the
University of Manchester, UK.} will be proposed in order to
investigate the link between parsec-scale and kiloparsec-scale
structures in this object.

\end{singlespace}
\bigskip

{\it Acknowledgements:This work was supported by the Spanish DGES
under grant AYA-2001-3490-C02 and Conselleria d'Empresa,
Universitat i Ci\`encia de la Generalitat Valenciana under project
GV2005/244. M.P. benefited from a postdoctoral fellowship in the
Max-Planck-Institut f\"ur Radioastronomie in Bonn and a Beca
Postdoctoral d'Excel$\cdot$l\`encia from Generalitat.}

\begin{small}

\end{small}

\end{document}